\documentclass[preprint,12pt,twocolumn, ]{elsarticle}

\usepackage{fixltx2e}  
\usepackage{makecell} 
\usepackage{multirow} 
\usepackage{hhline}   

\usepackage{amssymb}

\journal{Computers in Biology and Medicine}

\begin{document}


\twocolumn[{\begin{frontmatter}



\title{Cardiac Aging Detection Using Complexity Measures}


\author[a]{Karthi Balasubramanian}
\author[b]{Nithin Nagaraj}
\address[a]{Dept. of Electronics and Communication, Amrita School of Engineering, Amritapuri, Amrita Vishwa Vidyapeetham, Amrita University, India, Email: karthib77@gmail.com,  Ph: 91-476-2802709}
\address[b]{Consciousness Studies Programme, School of Humanities, National Institute of Advanced Studies,  Indian Institute of Science Campus, Bengaluru, India, Email: nithin@nias.iisc.ernet.in }

\begin{abstract}
As we age, our hearts undergo changes which result in reduction in complexity of physiological interactions between different control mechanisms. This results in a potential risk of cardiovascular diseases which are the number one cause of death globally. Since cardiac signals are nonstationary and nonlinear in nature, complexity measures are better suited to handle such data. In this study, non-invasive methods for detection of cardiac aging using complexity measures are explored. Lempel-Ziv (LZ) complexity, Approximate Entropy (ApEn) and Effort-to-Compress (ETC) measures are used to differentiate between healthy young and old subjects using heartbeat interval data.  We show that both LZ and ETC complexity measures are able to differentiate between young and old subjects with only 10 data samples while ApEn requires at least 15 data samples.
\end{abstract}


\end{frontmatter}
}]


\section{Introduction}
It is well known that functions of all physiological systems are greatly altered during the process of aging. Among these, the cardiovascular system has received prominent attention due to the high death rate attributed to heart related diseases. In fact, the World Health Organization (WHO) has labeled cardiovascular diseases as the number one cause of death globally \cite{deathCVD}. Hence, study of cardiac aging has been an important area in clinical medicine, and understanding heart rate patterns is an essential step in this study. In the 1980's, chaos theory began being used to investigate heart beat dynamics \cite{physiologyChaos80s1, physiologyChaos80s2, physiologyChaos80s3}. Initially, researchers had assumed that pathological systems were the ones that produce chaotic time series. They therefore tried using the concepts of nonlinear theory in  modeling pathologies like cardiac arrhythmias, which initiated the process of understanding the behaviour and dynamics of atrial or ventricular fibrillation \cite{heartbeatChaotic}. But contrary to earlier theories, it was found that cardiac arrhythmia doesn't display chaotic dynamics \cite{pathologyNotChaos}. It was indeed a surprise when further studies led to the discovery that cardiac chaos is not found in pathological systems but in the dynamics of normal sinus rhythm \cite{normalHeartChaos1, normalHeartChaos2, normalHeartChaos3}.

Goldberger, in his seminal work on heart beat dynamics \cite{heartbeatNonlinear1}, points out that even during bed rest, the heart rate in healthy individuals is not constant or periodic, but shows irregular patterns that are typically associated with a chaotic non-linear system. A detailed analysis of the beat-to-beat heart rate variability time series shows that there is no particular time scale for this behaviour and it is seen in all orders of magnitude (hours, minutes and seconds). This self-similar structure lends credence to the fact that a fractal feature of chaotic dynamics exists in these time series data. This has been further verified by Goldberger \cite{heartbeatNonlinear1} by performing spectral analysis on normal heart time series, which revealed a broad spectrum with a continuum of frequency components, indicating a high possibility of presence of chaos. Added to this, the phase space mapping of the heart rate time series produced strange attractors, not limit cycles, which clearly indicates the presence of underlying chaotic dynamics in the heart rate variability data.
It has been surmised that this chaotic behaviour is due to the fact that a healthy physiological condition is defined by complex interactions between multiple control mechanisms. These interactions are essential for the individual to adapt to the ever changing external environment. These highly adaptive and complex mechanisms lead to a very irregular firing rate of the pacemaker cells in the heart's sinus node, resulting in a chaotic heart rate variability \cite{heartbeatChaotic}.

Looking at this physiological response of healthy individuals, a natural question that arises is: `How long can this complex interaction continue? What happens as the person gets older?' This was the focus of study by Goldberger \textit{et al.} in \cite{complexityAging}. They point out that, during the process of aging, there occur two effects that affect the complexity of the interactions among the control mechanisms.
\begin{enumerate}
\item Progressive impairment of functional components.
\item Altering of nonlinear coupling between these components.
\end{enumerate}
Due to these effects, the physiological functions no longer show multitudes of variations but slowly settle down to a more regular nature. This results in a decrease of complexity in heart rate variability in healthy older individuals. This also explains why researchers are motivated to use complexity measures (such as approximate entropy, Lempel-Ziv complexity and other measures) to study cardiac aging, as such measures are known to aid the detection of the regularity inherent in the cardiac time series. From a clinical application perspective, it is desirable to have a complexity measure that can aid this detection in real-time (hence should be able to work with small number of samples), and also be robust to noise or missing data.

\section{Existing studies and their limitations}
Statistical measures like mean and variance can't adequately quantify complex chaotic-like behaviour. This is because systems with very different dynamics may have very similar means and variances. Hence Goldberger \textit{et al.} in \cite{cardiacAgingApEn} have used the following two measures to capture the complex nonlinear variability of the physiological processes.
\begin{enumerate}
\item Measurement of the dimension of the nonlinear system that describes the physiological data variations.
\item Calculation of ApEn (Approximate Entropy) of the physiological time series.
\end{enumerate}

Using beat-to-beat heart rate signals collected from 16 healthy young subjects and 18 healthy elderly subjects, they demonstrate that the complexity of cardiovascular dynamics reduces with aging.
Similar studies have also been done using fractal measures like `detrended fluctuation analysis' to quantify long range correlation properties in \cite{fractal1,fractal2} to show that complexity of heart beat variablity decreases with aging. Use of  multifractality, sample entropy and wavelet analysis has been explored in \cite{AgingMultifractality_sampleEntropy}. Nemati \textit{et al.} in \cite{agingTransferEntropy} use band-limited transfer entropy to characterize cardiac aging, with an added purpose of discovering the effect of respiration and blood pressure on complexity of heart rate variability  due to aging phenomenon.

All the aforementioned studies have used very long data sequences, running into 1000s of data samples for their analysis. Some methods that are used, namely `detrended fluctuation analysis' and approximate dimension require such large data sets to effectively quantify the complexity.

\par Looking at the fractal nature of the heart beat variability data, there is a high possibility that short data sequences would be enough to fully characterize the structure in the time series, provided the complexity measure can handle short sequences effectively. Takahashi in \cite {AgingHRV_conditionalentropy} has used conditional entropy and symbolic analysis with sequences of length 200 for differentiating between heart beat intervals of young and old subjects. We go a step ahead and use sequences shorter than that. Our goal is to compare the usefulness of Effort-to-Compress (ETC \cite{ETC}), Approximate Entropy (ApEn \cite{ApEn}), and Lempel-Ziv (LZ \cite{LZComplexity}) complexity measures in automatic identification of cardiac aging using very short length inter-beat time series data. It is conceivable that longer inter-beat time series data may not be practically available, or that the long sequence may be unreliable due to contamination by noise. In such instances, it is always beneficial to rely upon short length time series data, especially if it were to be useful in a practical setting.

In the next section, we briefly introduce the three complexity measures that we have used in our study. In section \ref{Experiment setup} we describe the experimental setup, followed by comparative analysis in section \ref{Comparative complexity analysis}.  We conclude with future research directions in section \ref{Conclusions}.

\section{Complexity Measures}
In this section, we introduce the reader to LZ, ApEn and ETC complexity measures.

\subsection{Lempel-Ziv complexity}
Lempel-Ziv complexity is a popular measure in the field of biomedical data characterization \cite{LZ_HRV, LZ_VentricularFibrillation}. To compute the Lempel-Ziv complexity, the given data (if numerical) has to be first converted to a symbolic sequence. This symbolic sequence is then parsed from left to right to identify the number of distinct patterns present in the sequence. This method of parsing is proposed in the seminal work on Lempel-Ziv complexity \cite{LZComplexity}. The very succinct description in \cite{LZfiniteDataSize} is reproduced here.
\\
Let $S\ =\ s_1s_2 \cdots s_n$ denote a symbolic sequence; $S(i,j)$ denote a substring of $S$ that starts at position $i$ and ends at position $j$; $V(S)$ denote the set of all substrings ($S(i,j)$ for $i\ =\ 1,2,\cdots n;$ and $j\ \geq \ i$). For example, let $S$ = $abc$, then $V(S)$ = \{$a,b,c,ab,bc,abc$\}. The parsing mechanism involves a left-to-right scan of the symbolic sequence $S$. Start with $i$ = 1 and $j$ = 1. A substring $S(i,j)$ is compared with all strings in $V(S(i,j-1))$ (Let V(S(1,0)) = \{\}, the empty set). If $S(i,j)$ is present in $V(S(1,j-1))$, then increase $j$ by 1 and repeat the process. If the substring is not present, then place a dot after $S(i,j)$ to indicate the end of a new component, set $i$ = $j$ + 1, increase $j$ by 1, and the process continues. This parsing procedure continues until $j\ = \ n$, where $n$ is the length of the symbolic sequence. For example, the sequence `$aacgacga$' is parsed as `$a.ac.g.acga$.'. By convention, a dot is placed after the last element of the symbolic sequence and the number of dots gives us the number of distinct words which is taken as the LZ complexity, denoted by $c(n)$. In this example, the number of distinct words (LZ complexity) is 4.

Since we may need to compare sequences of different lengths, a  normalized measure is proposed and is denoted by $C_{LZ}$ and expressed as :
\begin{equation}
C_{LZ} = (c(n)/n)log_{\alpha}n.
\end{equation}
where $\alpha$ denotes the number of unique symbols in the symbol set \cite{LZ_interpretation}.

\subsection{Approximate Entropy (ApEn)}
Approximate entropy is a complexity measure used to quantify regularity of time series, especially short and noisy sequences \cite{ApEn}. ApEn is a measure that monitors how much a set of patterns that are close together for a few observations, still retains its closeness on comparing the next few observations. Basically it checks for the  convergence and divergence of patterns to check the complexity of the given sequence. If neighbouring patterns retain the same closeness, then we infer it to be a more regular pattern, with a concomitant lower ApEn value. The measure been defined in \cite{ApEn} and we reproduce the definition here. Two input parameters, $m$ and $r$, must be initially chosen for the computation of the measure - $m$ being the length of the patterns we want to compare each time for closeness, and $r$ being a tolerance factor for the regularity of the two sets of patterns being compared. \\
Given a sequence $u$ of length N, we now define the complexity ApEn($m$,$r$,$N$) as follows.
\begin{itemize}
\item Form vector sequences $x(1)$ through $x(N-m+1)$ defined by $x(i)$ = [$u(i),u(i+1),...u(i+m-1)$], representing $m$ consecutive $u$ values, starting from the $i^{th}$ value.
\item Define the distance $d[x(i),x(j)]$ between vectors $x(i)$ and $x(j)$ as the maximum difference in their respective scalar components.

\item For each $i \leq N-m+1$, calculate the number of $j \leq N-m+1$ such that $d[x(i),x(j)] \leq r$ and call the number as $k(i)$.
\item For each $i \leq N-m+1$, calculate the parameters $C_i^m(r)$ = $k(i)/(N-m+1)$. These parameters measure, within a tolerance $r$, the regularity or frequency of patterns similar to given pattern of length $m$.
\item Define
\begin{equation}
\phi^m(r)= \frac{\sum_{i=1}^{N-m+1} ln \ C_i^m(r)}{N-m+1}
\end{equation}
\item Using this, ApEn complexity measure is defined as
\begin{equation}
\textrm{ApEn}(m,r,N) = \phi^m(r) - \phi^{m+1}(r)
\end{equation}
\end{itemize}

It has been shown in \cite{ApEn} that for $m$=1  and 2, values of $r$ between 0.1 to $0.25SD$(standard deviation) of the sequence provide good statistical validity of ApEn$(m,r,N)$. In our analysis, we use $m$=1 and $r$ = $0.25SD$(standard deviation) of the sequence.

\subsection{Effort-to-Compress (ETC) complexity}
Effort-to-Compress (ETC) is a recently proposed complexity measure that is based on the effort required by a lossless compression algorithm to compress a given sequence \cite{ETC}. The measure has been proposed using a lossless compression algorithm known as Non-sequential Recursive Pair Substitution (NSRPS). The algorithm for compressing a given sequence of symbols proceeds as follows. At the first iteration, that pair of symbols which has maximum number of occurrences is replaced with a new symbol. For example, the input sequence `$11010010$' is transformed into `$12202$' since the pair `$10$' has maximum number of occurrences compared to other pairs (`$00$', `$01$' and `$11$'). In the second iteration, `$12202$' is transformed to `$3202$' since `$12$' has maximum frequency (in fact all pairs are equally likely). The algorithm proceeds in this fashion until the length of the string is 1 or the string becomes a constant sequence (at which stage the entropy is zero and the algorithm halts). In this example, the algorithm transforms the input sequence `$11010010$' $\mapsto$ `$12202$' $\mapsto$ `$3202$' $\mapsto$ `$402$'  $\mapsto$ `$52$' $\mapsto$ `$6$'.

The ETC measure is defined as $N$, the number of iterations required for the input sequence to be transformed to a constant sequence through the usage of NSRPS algorithm.
$N$ is an integer that varies between 0 and $L-1$, where $L$ stands for the length of the input symbolic sequence. The normalized version of the measure is given by:
 $\frac{N}{L-1}$ (Note: $0 \leq \frac{N}{L-1} \leq 1$).

\section{Experimental setup} \label{Experiment setup}
Data used for our experiment was obtained from  `Physionet: Fantasia database' \cite {physionet} and has been described in \cite{fractal2}. Beat-to-beat heart rate signals from two groups of healthy adults: twenty young (age 21-34) and twenty old (age 68-81) with 10 males and females in each category, were obtained. The subjects were studied while lying in a supine (lying on back) position and watching the Disney movie `Fantasia' to maintain wakefulness. The ECG data was sampled at 250 Hz and the obtained samples were used for analysis. For further details of the experimental setup, please refer to \cite{fractal2}.

In our study, we take short length samples from random time instances and analyze them using the three complexity measures (described in the previous section) to gauge their efficiency in distinguishing between the two groups of data.

\section{Comparative complexity analysis} \label{Comparative complexity analysis}
In this section, we take a closer look at the performances of each of the measures. We are interested in finding out the minimum number of samples required for correct classification.
\subsection{Analysis procedure}
\begin{itemize}
\item From each of the twenty young and old data sets, choose consecutive $L$ number of samples from a random location.
\item Calculate the ApEn, LZ and ETC complexity measures for the chosen $L$ length data set.
\item For statistical accuracy, 50 such locations are randomly chosen and 50 complexity values for each of the measures are calculated.
\item The complexity assigned to each of the measure for a sequence of length $L$ is the average of all the 50 values calculated.
\item Thus for each complexity measure, we obtain 20 complexity values for the young subjects and 20 complexity values for the old subjects.
\item Using these values as samples representing the young and old populations, a two-sample t-test is performed for each complexity measure.
\item The results of the t-test are analyzed to check if the mean complexity value of the beat-to-beat interval of the young subjects is significantly greater than the mean complexity value of the beat-to-beat interval of the old subjects.
\item The entire process is repeated multiple times with different values of $L$ and the minimum length at which each complexity measure is able to successfully classify is determined.
\end{itemize}
\subsection{Results summary} \label{Results summary}
As per the procedure outlined above, complexity measures were calculated for data of different lengths. The continuous valued beat-to-beat interval data was quantized using using 4 bins and again using 8 bins and complexity measures were calculated. Two-sample t-test was performed on the complexity values for various lengths and the results are shown in Tables \ref{tab:t-test_4bins} and \ref{tab:t-test_8bins} respectively.

\begin{table*}[t]
\begin{minipage}{\textwidth}
\begin{center}
\caption{Student's t-test results for ApEn, LZ and ETC complexity analysis to distinguish between beat-to-beat intervals of healthy young and old subjects using 4 bins.}
\label{tab:t-test_4bins}
	\begin{tabular}{|c||c||c|c|c|c|c|}
  		\hline
  		\textbf{Length} &\textbf{Complexity} & 		        \textbf{Mean(Old)} & \textbf{Mean(Young)}  & \textbf{t-value} & \textbf{df} & \textbf{p}\\

		\hline
  		\multirow{2}{*}{L=20}
  		& ApEn & 0.718 & 0.769 & -3.29 & 30 & 0.001 \\
  		\hhline{~------}
  		& LZ & 0.871 & 0.928 & -2.76 & 30 & 0.005 \\
     	\hhline{~------}
  		&ETC & 0.712 & 0.729 & -1.86 & 28 & 0.037 \\   	
  		\hline
  		
  		\multirow{2}{*}{L=15}
 		& ApEn & 0.588 & 0.621 & -2.24 & 27 & 0.017 \\
  		\hhline{~------}
  		& LZ & 0.905 & 0.948 & -2.36 & 34 & 0.012 \\
     	\hhline{~------}
  		&ETC & 0.771 & 0.778 & -0.87 & 27 & 0.195 \\   	
  		\hline
  		
  		\multirow{2}{*}{L=10}
  		& ApEn & 0.358 & 0.385 & -1.41 & 38 & 0.083 \\
  		\hhline{~------}
  		& LZ & 0.958 & 0.963 & -0.38 & 38 & 0.352 \\
     	\hhline{~------}
  		&ETC & 0.869 & 0.871 & -0.15 & 38 & 0.439 \\   	
  		\hline  		
\end{tabular}

\bigskip

\caption{Student's t-test results for ApEn, LZ and ETC complexity analysis to distinguish between beat-to-beat intervals of healthy young and old subjects using 8 bins.}
\label{tab:t-test_8bins}
	\begin{tabular}{|c||c||c|c|c|c|c|}
  		\hline
  		\textbf{Length} &\textbf{Complexity} & 		        \textbf{Mean(Old)} & \textbf{Mean(Young)}  & \textbf{t-value} & \textbf{df} & \textbf{p}\\

		\hline
  		\multirow{2}{*}{L=15}
  		& ApEn & 0.588 & 0.621 & -2.24 & 27 & 0.017 \\
  		\hhline{~------}
  		& LZ & 0.795 & 0.832 & -3.09 & 32 & 0.004 \\
     	\hhline{~------}
  		&ETC & 0.883 & 0.909 & -3.01 & 29 & 0.005 \\   	
  		\hline
  		
  		\multirow{2}{*}{L=10}
  		& ApEn & 0.358 & 0.385 & -1.41 & 38 & 0.083 \\
  		\hhline{~------}
  		& LZ & 0.788 & 0.809 & -2.10 & 38 & 0.043 \\
     	\hhline{~------}
  		&ETC & 0.933 & 0.950 & -2.30 & 38 & 0.027 \\   	
  		\hline
  		
  		\multirow{2}{*}{L=8}
  		& ApEn & 0.305 & 0.309 & -0.26 & 38 & 0.797 \\
  		\hhline{~------}
  		& LZ & 0.773 & 0.790 & -1.89 & 38 & 0.067 \\
     	\hhline{~------}
  		&ETC & 0.956 & 0.962 & -1.25 & 38 & 0.217 \\   	
  		\hline  		
\end{tabular}

\end{center}
\end{minipage}
\end{table*}
The t-test results for analysis using 4 bins may be summarized as follows:
\begin{itemize}
\item The mean ApEn complexity of the beat-to-beat interval of old subjects is significantly less than that of young subjects for data lengths of 20 (\textit{t}\textsubscript{\bf {30}} = -3.29,  \textit{p} = 0.001) and 15 (\textit{t}\textsubscript{\bf {27}} = -2.24,  \textit{p} = 0.017) while it is not significantly less (\textit{t}\textsubscript{\bf {38}} = -1.41,  \textit{p} = 0.083) for data length of 10.

\item The mean LZ complexity of the beat-to-beat interval of old subjects is significantly less than that of young subjects for data lengths of 20 (\textit{t}\textsubscript{\bf {30}} = -2.76,  \textit{p} = 0.005) and 15 (\textit{t}\textsubscript{\bf {34}} = -2.36,  \textit{p} = 0.012) while it is not significantly less (\textit{t}\textsubscript{\bf {38}} = -0.38,  \textit{p} = 0.352) for data length of 10.

\item The mean ETC complexity of the beat-to-beat interval of old subjects is significantly less than that of young subjects for a data length of 20 (\textit{t}\textsubscript{\bf {28}} = -1.86 \textit{p} = 0.037), while it is not significantly less for data lengths of 15 (\textit{t}\textsubscript{\bf {27}} = -0.87,  \textit{p} = 0.195) and 10 (\textit{t}\textsubscript{\bf {38}} = -0.15,  \textit{p} = 0.439).
\end{itemize}
%
%
%
%


The t-test results for analysis using 8 bins may be summarized as follows:

\begin{itemize}
\item The mean ApEn complexity of the beat-to-beat interval of old subjects is significantly less (\textit{t}\textsubscript{\bf {27}} = -2.24,  \textit{p} = 0.017) than that of young subjects for a data length of 15  while it is not significantly less  for data lengths of 10 (\textit{t}\textsubscript{\bf {38}} = -1.41,  \textit{p} = 0.083) and 8 (\textit{t}\textsubscript{\bf {38}} = -0.26,  \textit{p} = 0.797).

\item The mean LZ complexity of the beat-to-beat interval of old subjects is significantly less than that of young subjects for data lengths of 15 (\textit{t}\textsubscript{\bf {32}} = -3.09,  \textit{p} = 0.004) and 10 (\textit{t}\textsubscript{\bf {38}} = -2.10,  \textit{p} = 0.043) while it is not significantly less (\textit{t}\textsubscript{\bf {38}} = -1.89,  \textit{p} = 0.067) for a data length of 8.

\item The mean ETC complexity of the beat-to-beat interval of old subjects is significantly less than that of young subjects for data lengths of 15 (\textit{t}\textsubscript{\bf {29}} = -3.01,  \textit{p} = 0.005) and 10 (\textit{t}\textsubscript{\bf {38}} = -2.30,  \textit{p} = 0.027) while it is not significantly less (\textit{t}\textsubscript{\bf {38}} = -1.25,  \textit{p} = 0.217) for a data length of 8.

\end{itemize}
Tables \ref{tab:heartbeat4bins} and \ref{tab:heartbeat8bins} summarize the ability of the complexity measures to classify beat-to-beat intervals of old and young subjects, using 4 bins and 8 bins respectively.
%

\begin{table}[!h]
\begin{center}

\caption{Ability of complexity measures for classifying the beat-to-beat interval of young and old subjects for different data lengths using 4 bins.}
\label{tab:heartbeat4bins}
\begin{tabular}{|c||c|c|c|c|}
  \hline
  \textbf{\thead{Complexity\\measure}} & \textbf{L=20} & \textbf{L=15} & \textbf{L=10}\\

     \hline
  ApEn     & Yes & Yes & No \\
     \hline
  LZ       & Yes & Yes & No \\
     \hline
  ETC      & Yes & No & No \\
     \hline
\end{tabular}
\end{center}
\end{table}

\begin{table}[!h]
\begin{center}
\caption{Ability of complexity measures for classifying the beat-to-beat interval of young and old subjects for different data lengths using 8 bins.}
\label{tab:heartbeat8bins}
\begin{tabular}{|c||c|c|c|c|}
  \hline
  \textbf{\thead{Complexity\\measure}} & \textbf{L=15} & \textbf{L=10} & \textbf{L=8}\\

     \hline
  ApEn     & Yes & No & No \\
     \hline
  LZ       & Yes & Yes & No \\
     \hline
  ETC      & Yes & Yes & No \\
     \hline
\end{tabular}
\end{center}
\end{table}
%

%
\section{Conclusions and Future Research Work} \label{Conclusions}
Based on our study on the experimental data, at a 5\% significance level (overall error rate) for the statistical test, there is sufficient evidence to conclude that:
\begin{itemize}
\item For data analyzed using 4 bins, ApEn and LZ complexity measures are able to distinguish between beat-to-beat intervals of  young and old subjects for lengths of 15 or more, while ETC complexity measure is able to do so for lengths of 20 and higher.
\item For data analyzed using 8 bins, both LZ and ETC complexity measures are able to distinguish between beat-to-beat intervals of  young and old subjects for lengths as short as 10 data samples, while ApEn is able to do so only for lengths greater than 15.
\end{itemize}
\par For future research work, it is imperative that we study the effect of noise (which is unavoidable in real-life applications) and missing-data on the complexity measures and how it might impact the discrimination between cardiac signals of young and old subjects. This could be done by adding different levels of noise on the existing data-sets and also removing parts of data to simulate missing-data problem. This is also important from a clinical application perspective.


\section{Acknowledgment}
The authors would like to acknowledge Gayathri R Prabhu (Indian institute of Technology (IIT), Chennai), Sutirth Dey (Indian institute of Science Educational and Research (IISER), Pune) and Sriram Devanathan (Amrita University) in this work. We also thank Del Marshall (Amrita University) for valuable suggestions​ to improve the manuscript.







\section{\textbf{References}}
\bibliographystyle{unsrt}   
\bibliography{ReferencesCardiacAging}

\begin{thebibliography}{10}

\bibitem{deathCVD}
Shanthi Mendis, Pekka Puska, Bo~Norrving, et~al.
\newblock Global atlas on cardiovascular disease prevention and control.
\newblock World Health Organization, 2011.

\bibitem{physiologyChaos80s1}
Ary~L Goldberger, Valmik Bhargava, Bruce~J West, and Arnold~J Mandell.
\newblock On a mechanism of cardiac electrical stability. the fractal
  hypothesis.
\newblock {\em Biophysical journal}, 48(3):525, 1985.

\bibitem{physiologyChaos80s2}
Ary~L Goldberger and Bruce~J West.
\newblock Fractals in physiology and medicine.
\newblock {\em The Yale journal of biology and medicine}, 60(5):421, 1987.

\bibitem{physiologyChaos80s3}
AL~Goldberger, DR~Rigney, J~Mietus, EM~Antman, and S~Greenwald.
\newblock Nonlinear dynamics in sudden cardiac death syndrome: heartrate
  oscillations and bifurcations.
\newblock {\em Cellular and Molecular Life Sciences}, 44(11):983--987, 1988.

\bibitem{heartbeatChaotic}
Ary~L Goldberger.
\newblock Is the normal heartbeat chaotic or homeostatic?
\newblock {\em Physiology}, 6(2):87--91, 1991.

\bibitem{pathologyNotChaos}
Ary~L Goldberger and David~R Rigney.
\newblock Nonlinear dynamics at the bedside.
\newblock In {\em Theory of Heart}, pages 583--605. Springer, 1991.

\bibitem{normalHeartChaos1}
Ary~L Goldberger and R~Rigney, David.
\newblock Chaos and fractals in human physiology.
\newblock {\em Scientific American}, 262:42--49, 1990.

\bibitem{normalHeartChaos2}
R~R{\"o}ssler, F~Gotz, and OE~R{\"o}ssler.
\newblock Chaos in endocrinology.
\newblock {\em Biophys. J}, 25(2):216a, 1979.

\bibitem{normalHeartChaos3}
Niels Wessel, Maik Riedl, and J{\"u}rgen Kurths.
\newblock Is the normal heart rate “chaotic” due to respiration?
\newblock {\em Chaos: An Interdisciplinary Journal of Nonlinear Science},
  19(2):028508, 2009.

\bibitem{heartbeatNonlinear1}
Ari~L Goldberger.
\newblock Is the normal heartbeat chaotic or homeostatic?
\newblock {\em News Physiol Sci}, 6:87--91, 1991.

\bibitem{complexityAging}
Lewis~A Lipsitz and Ary~L Goldberger.
\newblock Loss of complexity and aging.
\newblock {\em Jama}, 267(13):1806--1809, 1992.

\bibitem{cardiacAgingApEn}
DT~Kaplan, MI~Furman, SM~Pincus, SM~Ryan, LA~Lipsitz, and AL~Goldberger.
\newblock Aging and the complexity of cardiovascular dynamics.
\newblock {\em Biophysical Journal}, 59(4):945--949, 1991.

\bibitem{fractal1}
Robb~W Glenny, H~Thomas Robertson, Stanley Yamashiro, and James~B
  Bassingthwaighte.
\newblock Applications of fractal analysis to physiology.
\newblock {\em Journal of Applied Physiology}, 70(6):2351--2367, 1991.

\bibitem{fractal2}
Nikhil Iyengar, C~K Peng, Raymond Morin, A~L Goldberger, and Lewis~A Lipsitz.
\newblock Age-related alterations in the fractal scaling of cardiac interbeat
  interval dynamics.
\newblock {\em American Journal of Physiology-Regulatory, Integrative and
  Comparative Physiology}, 271(4):R1078--R1084, 1996.

\bibitem{AgingMultifractality_sampleEntropy}
Anne Humeau, Fran{\c{c}}ois Chapeau-Blondeau, David Rousseau, Pascal Rousseau,
  Wojciech Trzepizur, and Pierre Abraham.
\newblock Multifractality, sample entropy, and wavelet analyses for age-related
  changes in the peripheral cardiovascular system: preliminary results.
\newblock {\em Medical Physics}, 35(2):717--723, 2008.

\bibitem{agingTransferEntropy}
Shamim Nemati, Bradley~A Edwards, Joon Lee, Benjamin Pittman-Polletta, James~P
  Butler, and Atul Malhotra.
\newblock Respiration and heart rate complexity: effects of age and gender
  assessed by band-limited transfer entropy.
\newblock {\em Respiratory physiology \& neurobiology}, 189(1):27--33, 2013.

\bibitem{AgingHRV_conditionalentropy}
Anielle~CM Takahashi, Alberto Porta, Ruth~C Melo, Robison~J Quit{\'e}rio, Ester
  da~Silva, Audrey Borghi-Silva, Eleonora Tobaldini, Nicola Montano, and
  Aparecida~M Catai.
\newblock Aging reduces complexity of heart rate variability assessed by
  conditional entropy and symbolic analysis.
\newblock {\em Internal and emergency medicine}, 7(3):229--235, 2012.

\bibitem{ETC}
Nithin Nagaraj, Karthi Balasubramanian, and Sutirth Dey.
\newblock A new complexity measure for time series analysis and classification.
\newblock {\em The European Physical Journal Special Topics},
  222(3-4):847--860, 2013.

\bibitem{ApEn}
Steve Pincus.
\newblock Approximate entropy ({A}p{E}n) as a complexity measure.
\newblock {\em Chaos: An Interdisciplinary Journal of Nonlinear Science},
  5(1):110--117, 1995.

\bibitem{LZComplexity}
Abraham Lempel and Jacob Ziv.
\newblock On the complexity of finite sequences.
\newblock {\em Information Theory, IEEE Transactions on}, 22(1):75--81, 1976.

\bibitem{LZ_HRV}
J~Goḿez-Pilar, GC~Guti{\'e}rrez-Tobal, D~Alvarez, F~del Campo, and R~Hornero.
\newblock Classification methods from heart rate variability to assist in
  {SAHS} diagnosis.
\newblock {\em XIII Mediterranean Conference on Medical and Biological
  Engineering and Computing 2013}, pages 1825--1828, 2014, Springer.

\bibitem{LZ_VentricularFibrillation}
Deling Xia, Qingfang Meng, Yuehui Chen, and Zaiguo Zhang.
\newblock Classification of ventricular tachycardia and fibrillation based on
  the lempel-ziv complexity and {EMD}.
\newblock {\em Intelligent Computing in Bioinformatics}, pages 322--329, 2014,
  Springer.

\bibitem{LZfiniteDataSize}
Jing Hu, Jianbo Gao, and Jose~Carlos Principe.
\newblock Analysis of biomedical signals by the {L}empel-{Z}iv complexity: the
  effect of finite data size.
\newblock {\em Biomedical Engineering, IEEE Transactions on},
  53(12):2606--2609, 2006.

\bibitem{LZ_interpretation}
Mateo Aboy, Roberto Hornero, Daniel Ab{\'a}solo, and Daniel {\'A}lvarez.
\newblock Interpretation of the {L}empel-{Z}iv complexity measure in the
  context of biomedical signal analysis.
\newblock {\em Biomedical Engineering, IEEE Transactions on},
  53(11):2282--2288, 2006.

\bibitem{physionet}
Ary~L Goldberger, Luis~AN Amaral, Leon Glass, Jeffrey~M Hausdorff, Plamen~Ch
  Ivanov, Roger~G Mark, Joseph~E Mietus, George~B Moody, Chung-Kang Peng, and
  H~Eugene Stanley.
\newblock Physiobank, physiotoolkit, and physionet components of a new research
  resource for complex physiologic signals.
\newblock {\em Circulation}, 101(23):e215--e220, 2000.

\end{thebibliography}

\end{document}